\documentclass[11pt,a4wide]{article}
\usepackage{graphicx}
\begin{document}
\begin{center}
{\Large\bf Resonant normal form and asymptotic normal form behavior\\
in magnetic bottle Hamiltonians} \vskip 0.5cm {C. Efthymiopoulos,
M. Harsoula and G. Contopoulos} \vskip 0.5cm
Research Center for Astronomy and Applied Mathematics\\
Academy of Athens, Soranou Efessiou 4, 115 27 Athens, Greece \vskip
1cm \noindent {\bf Keywords: } 
Normal forms; Magnetic bottle; Magnetic moment; Resonance; Chaos\vskip 1cm
\end{center}

\noindent{\small {\bf Abstract:} 
We consider normal forms in `magnetic bottle' type Hamiltonians 
of the form $H=\frac{1}{2}(\rho^2_\rho+\omega^2_1\rho^2)
+\frac{1}{2}p^2_z+hot$ (second frequency $\omega_2$ equal to zero in 
the lowest order). Our main results are: i) a novel method to construct 
the normal form in cases of resonance, and ii) a study of the asymptotic 
behavior of both the non-resonant and the resonant series. We find that, 
if we truncate the normal form series at order $r$, the series remainder 
in both constructions decreases with increasing $r$ down to a minimum, 
and then it increases with $r$. The computed minimum remainder turns to 
be exponentially small in $\frac{1}{\Delta E}$, where $\Delta E$ is the 
mirror oscillation energy, while the optimal order scales as an inverse 
power of $\Delta E$. We estimate numerically the exponents associated 
with the optimal order and the remainder's exponential asymptotic behavior. 
In the resonant case, our novel method allows to compute a `quasi-integral' 
(i.e. truncated formal integral) valid both for each particular resonance 
as well as away from all resonances. We applied these results to a specific 
magnetic bottle Hamiltonian. The non resonant normal form yields 
theorerical invariant curves on a surface of section which fit well the 
empirical curves away from resonances. On the other hand the resonant 
normal form fits very well both the invariant curves inside the islands 
of a particular resonance as well as the non-resonant invariant curves.
Finally, we discuss how normal forms allow to compute a critical 
threshold for the onset of global chaos in the magnetic bottle. }
\section{Introduction}

`Magnetic bottle' type nonlinear Hamiltonian dynamical systems
appear in various areas of physics and astronomy. Examples are
plasma confinement machines, ion traps and charged particle
measuring devices, planetary magnetospheres leading, e.g., to the
formation of radiation belts, magnetic reconnection, magnetic
bottles in the solar corona, etc. (see, for example, Dendy 1993,
Gurnett and Bhattacharjee 2005, and references therein).

It is well known that in such configurations there exist both
regular and chaotic particle orbits. The regular orbits are of
oscillatory nature, i.e., a gyration around the magnetic field lines
combined with a `mirror' oscillation along the field lines. A
bouncing of the particles appears when the magnetic field lines
converge towards a preferential direction (see e.g. Jackson 1962). A
magnetic bottle is formed when there are two distinct domains along
the field lines in which we have such a convergence. The particles
are reflected as they approach the two `necks' of the bottle. The
mirror frequency is of order $\omega_z = O(|V_\perp^2 (\partial^2
B_{\perp}/\partial\rho\partial z)/B_z)^{1/2}|)$, where $V_\perp$ is
the gyration velocity, $B_z$, $B_\perp$ are the measures of the
magnetic field along and accross the preferential direction (denoted
by $z$) respectively, and $\rho\perp z$. One typically has
$\omega_z<<\omega_c$, where $\omega_c = B_zq/m$ is the gyrofrequency
of a particle of charge $q$ and mass $m$.

A basic form of adiabatic theory (see, for example, Jackson 1962, or
Lichtenberg and Lieberman 1992), describes mirror oscillations as a
consequence of the preservation of a so-called `adiabatic invariant'. 
To lowest order, this corresponds to the particle's magnetic dipole 
moment $\mu = qV_\perp^2/(2\omega_c)$. The determination of higher 
order (in powers of $\mu$) adiabatic invariants is a classical problem 
of dynamics. Several methods to deal with this problem are discussed 
in Kruskal (1962), Northrop (1963), Contopoulos (1965), Dragt (1965), 
Arnold et al. (1988), Lichtenberg and Lieberman (1992) and Benettin 
et al. (1999).

For magnetic bottle Hamiltonians quite efficient methods can be
derived in the context of canonical perturbation theory. Such
methods have been proposed by Contopoulos and Vlahos (1975), and
Dragt and Finn (1979). In the canonical context, one starts from 
the basic Hamiltonian
\begin{equation}\label{ham1}
H={1\over 2m}\left(\mathbf{p}- q\mathbf{A}\right)^2
\end{equation}
where $\mathbf{A}$ is the vector potential corresponding to the
magnetic field $\mathbf{B}=\nabla\times\mathbf{A}$ and $\mathbf{p}$
are generalized momenta conjugate to the position variables.
Assuming, in the simplest case, axisymmetry around the z-axis, the
Hamiltonian in cylindrical co-ordinates $(\rho,z)$ takes the form
\begin{equation}\label{ham2}
H={p_\rho^2\over 2}+{1\over 2}\omega_1^2 \rho^2 + {p_z^2\over 2}+...
\end{equation}
with $\omega_1=\omega_c$ for particles gyrating around the z-axis. 
Note that the equations of motion arising from the quadratic 
terms in the Hamiltonian (\ref{ham2}) represent a case of so-called 
`nilpotent' linearization, since one of the eigenvalues of the 
matrix of the linearized system of equations is equal to zero. There 
is a variety of methods for constructing a normal form for such 
systems (see, for example, Meyer (1984), Cushman and Sanders (1986), 
Elphick (1988), Baider and Sanders (1991), as well as Sanders et al. 
(2007) for a review). On the other hand, the canonical approach leads 
to a definition of action-angle variables for the magnetic bottle, 
admitting straightforward physical interpretation.  In particular, 
if we define the pair of action-angle variables $(J_1,\phi_1)$ via
\begin{equation}\label{adact}
\rho=\sqrt{2J_1\over\omega_1}\sin\phi_1,~~~ p_\rho=\sqrt{2\omega_1
J_1}\cos\phi_1~~,
\end{equation}
the quantity $J_1$ is proportional to the magnetic dipole moment, 
i.e., $J_1=mV_\perp^2/(2\omega_c)$ $=(m/q)\mu$. After some steps 
of perturbation theory, we can then define new canonical 
variables $(\theta_1,I_1,\zeta , P_{\zeta})$, which are near-identity
transformations of the old canonical variables $(\phi_1,J_1,z,p_z)$,
so that the Hamiltonian in the new variables takes the form:
\begin{equation}\label{hamnf}
H(\theta_1,I_1,\zeta ,P_{\zeta})=Z(I_1,\zeta
,P_{\zeta})+R(\theta_1,I_1,\zeta ,P_{\zeta})~~.
\end{equation}
The function $Z$, called normal form, has the form
\begin{equation}\label{hamnf}
Z(I_1,\zeta ,P_{\zeta})=\omega_1I_1 + {1\over 2}(P_{\zeta}^2 +
\omega_2^2(I_1)\zeta^2)+...
\end{equation}
The action $I_1$ would be an exact integral of the Hamiltonian flow under
$Z$ alone. On the other hand, the function $R$, called `remainder', depends
on the angle $\theta_1$, thus it introduces some time variations of $I_1$
under the complete Hamiltonian flow of ({\ref{hamnf}). Nevertheless, one
typically has that $|R|<<|Z|$, implying that the effect of the remainder
on dynamics is small. Hence, $I_1$ represents a quasi-integral of motion,
i.e. a high-order adiabatic invariant, while the mirror frequency 
$\omega_2$ (assumed small) is expressed by this theory as a 
function of $I_1$.

The convergence behavior of the magnetic bottle normal form series 
at high normalization orders has not yet been fully explored. A systematic 
study of the convergence is presented in Engel et al. (1995). These authors
considered a polynomial magnetic bottle model. Then, they computed a
polynomial form of a `quasi-integral' up to degree 14 in the 
canonical variables. From the numerical data, they distinguish 
three types of behavior, i.e., i) convergence, ii) non-convergence, 
and (iii) `pseudo-convergence', depending on the behavior of the numerical 
variations of their quasi-integral at various normalization orders up to 
order 14. Here, we extend this analysis to much higher orders, and provide 
a numerical estimate of the asymptotic behavior of the series based on the 
size of the remainder of the normal form. In agreement with basic theory, 
we observe numerically that the only existing behavior is `pseudo-convergence', 
i.e., the series exhibit always an asymptotic behavior. This means that 
an `optimal' normalization order $r_{opt}$ can be identified, up to which 
the remainder decreases in size, yielding the impression that the normalization 
is convergent, while, beyond the optimal order, the size of the remainder 
increases with the normalization order, thus the normalization turns 
always to be divergent. We also provide evidence that the crucial small 
quantity which enters in all asymptotic estimates is the energy $\Delta E$ 
of the mirror oscillations. Asymptotically, one has the estimates 
$r_{opt}={\cal O}(\Delta E^{-\alpha})$, for the optimal order, 
and $||R||_{opt} ={\cal O}(\exp(-1/\Delta E^\gamma))$ for the size 
of the remainder at the optimal order, with exponents $\alpha$ and 
$\gamma$ specified numerically in section 3 below. 
We note that theoretical exponential estimates on adiabatic
invariants in nonlinear modulated oscillators are discussed in
Neishtadt (1981,1984) and Benettin and Sempio (1994) (for the case
of linear oscillators, see Howard (1970) and references therein).

The second main result in the present paper regards the construction 
of a normal form in magnetic bottle Hamiltonians in a case not covered 
by the usual theory, namely the case of {\it resonances}. Resonances 
appear whenever the condition $\omega_2/\omega_1=m_2/m_1$ is satisfied 
for non-zero integers $m_1,m_2$.  Such values are marked by the bifurcation 
of new periodic orbits (in pairs stable-unstable) from a so-called
`central' (equatorial) periodic orbit ($z=0$). The most important resonances 
are of the form $\omega_1-n\omega_2=0$, with $n$ integer. Resonances 
of lower and lower order appear by increasing the energy. The 
appearance of the lowest resonances 1:4, 1:3, 1:2, marks an overall 
qualitative change of the phase space structure leading eventually 
to the onset of global chaos.

The usual (non-resonant) normal form can predict the values of 
$I_1$ when new resonances bifurcate, as well as the distance of the 
resonances from the central orbit as $I_1$ increases (Contopoulos and 
Vlahos 1975). Nevertheless, it cannot describe the structure of the 
phase space near resonances. Here, precisely, we propose a method of 
construction of a resonant normal form for magnetic bottle 
Hamiltonians, which is applicable both for resonant orbits of one 
(at a time) specific resonance as well as for the non-resonant orbits 
in its neighborhood. Our method can be viewed as a combination of two 
recently introduced techniques: these are i) `detuning' (see Pucacco 
et al. 2008), and ii) `book-keeping' (see Efthymiopoulos 2008, 2012). 
Both techniques reflect ways to optimize the formal treatment of 
various small quantities appearing in the formal series. The main 
difference between the usual non-resonant construction and the hereby 
proposed resonant construction is the following: In the case of non-resonant 
series, we select as small parameter {\it either} the frequency $\omega_2$
or the distance (in phase space) from the central orbit (see
Contopoulos (1965) for a detailed comparison of the two approaches).
In the resonant series, however, we {\it simultaneously} treat the
distance from the central orbit and the small frequencies as small
parameters. We note, finally, that the case presently dealt with is
quite distinct from the case of resonance in models with a periodic
space modulation of the magnetic field (as e.g. in Dunnett et al.
1968, McNamara 1978).

Implementing our algorithm we compute high order resonant 
quasi-integrals, and check their degree of accuracy in comparison with 
the invariant curves found numerically on the domain of regular motion,
using as reference the same model as Engel et al. (1995). In 
the same way as for the non-resonant normal form, we also here 
examine the asymptotic behavior of the resonant formal series. We
demonstrate that in this case as well there hold exponential
estimates for the dependence of the optimal normalization order, as
well as the size of the optimal remainder, on the mirror oscillation
energy $\Delta E$. As a final outcome, we use the magnetic 
bottle normal forms in order to analytically compute the 
critical energy beyond which the central periodic orbit becomes 
unstable. This determines the energy where we have the onset 
of global chaos in the magnetic bottle.

The paper is organized as follows. Section 2 briefly describes 
some features of the reference model (Engel et al. 1995) used in our study,  
for the paper's self-containedness. Section 3 describes the algorithm 
of computation of the non-resonant normal form. We emphasize that 
this is not a new method but essentially the same as in Dragt and Finn 
(1979) and Engel et al (1995). Also, here as well we employ the method 
of Lie series which is quite convenient for performing near-identity 
canonical transformations (Hori (1966), Deprit (1969)). However, 
we present the specific algorithmic steps in greater detail than in these 
earlier papers, in order to set the context and introduce the notation 
of the `book-keeping' technique (Efthymiopoulos 2012). The main new result 
in this section regards the numerical study of the asymptotic 
properties of the non-resonant series and the determination of the 
associated exponents entering in exponential estimates of the size 
of the optimal remainder. These are found by extending all normal form 
computations up to a high order. Section 4 describes the novel computation 
of the resonant normal form construction for magnetic bottle 
Hamiltonians. Here, also, we study numerically the normal form's 
asymptotic behavior by reaching sufficiently high normalization 
orders. Finally, we compute approximately the threshold for the 
onset of global chaos using normal forms. Section 5 summarizes 
our conclusions.

\section{Hamiltonian model}
We consider the same axisymmetric magnetic bottle model as in Engel 
et al. (1995). The magnetic field is given by 
$\mathbf{B}=\mathbf{\nabla}\times\mathbf{A}$, 
where $\mathbf{A}$ is the vector potential given in cylindrical coordinates 
by $\mathbf{A}\equiv$ $(A_\rho,A_\phi,A_z)=$ $(0,A_\phi,0)$ with 
\begin{equation}\label{aphi}
A_{\phi}={B_0\over 2}
\left(\rho -\beta_1({1\over 8}\rho^3 - {1\over 2}\rho z^2)\right)~~.
\end{equation}
$B_0$ is the value of the homogeneous (along $z$) magnetic field component 
and $\beta_1$ measures the strength of the (octupole) component causing the 
mirroring effect. In the physical context, $\beta_1$ represents a small 
parameter. The equations of motion for a particle of mass $m=1$ and 
charge $q=1$ are derived from the Hamiltonian
\begin{equation}\label{hamgen}
H={1\over 2}\left(\mathbf{p}-\mathbf{A}\right)^2=
{p_z^2\over 2}+{p_\rho^2\over 2}+{p_\phi^2\over 2\rho^2}
-{A_\phi p_\phi\over\rho} + {A_\phi^2\over 2}~~. 
\end{equation}
\begin{figure}[h]
\centering
\includegraphics[scale=0.27]{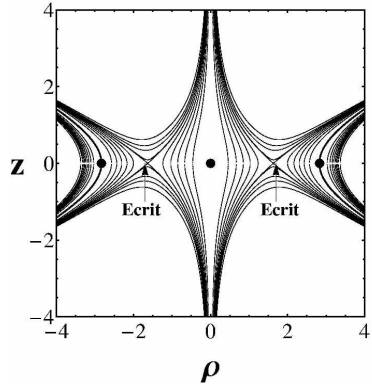}
\caption{\small The family of equipotential curves or curves of zero velocity
(CZV), $E$=$V(\rho,z)=A_\phi^2/2$, for various values of $E$. $E_{crit}$=$0.
\overline{592}$ is the value of $V(r,0)$ at its local maximum, and the dots 
indicate the local minima $E=0$.}
\label{isopot}
\end{figure}
Since $\phi$ is ignorable in (\ref{hamgen}), $p_{\phi}$ is an integral of 
motion. The system can be considered as of two degrees of freedom, with 
`effective potential':
\begin{equation}\label{veff}
V_{eff}= {p_\phi^2\over 2\rho^2}-{A_\phi(\rho,z) p_\phi\over\rho} 
+ {A_\phi^2(\rho,z)\over 2}~~.
\end{equation}
Orbits starting with $z=0$, $p_z=\dot{z}=0$ remain always on the equatorial 
plane $z=0$. For such orbits, the equations $\dot{\rho}=p_{\rho}=0$ and 
$\dot{p_\rho}=-\partial V_{eff}(\rho,z=0)/\partial\rho=0$ define two types 
of equilibria for any fixed radius $\rho$:\\ 
\\
i) For $p_{\phi}<0$ the equilibrium solution is 
$\rho=\rho_c=[-p_\phi\/\partial A_\phi/\partial\rho)_{\rho=\rho_c,z=0}]^{1/2}$.
This yields a circular equatorial orbit surrounding the central axis, 
with gyration frequency $\omega_c={p_\phi/\rho_c^2}-A_\phi(\rho_c,0)/\rho_c<0$.
Nearby orbits can be studied by means of the epicyclic approximation. 
Setting $\xi=\rho-\rho_c$ and expanding the Hamiltonian around the 
circular solution we arrive at:
\begin{equation}\label{hamepi}
H=\mbox{const}+{p_z^2\over 2} +{p_\xi^2\over 2}
+{1\over 2}\kappa^2\xi^2 +H_1
\end{equation}
where $\kappa^2 = 3p_\phi^2/\rho_c^4+2p_\phi A'_\phi/\rho_c^2
-2p_\phi A_\phi/\rho_c^3-p_\phi A''_\phi/\rho +A''_\phi A_\phi + (A'_\phi)^2$
(derivatives are with respect to $\rho$, evaluated at $\rho=\rho_c,z=0$).
It is easy to check that in $H_1$ the lowest order terms quadratic in $z$ 
are either of order $O(\beta_1z^2)$ or $O(\xi z^2)$, hence small. Thus, 
the Hamiltonian (\ref{hamepi}) is of the general form (\ref{ham2}).  \\
\\
ii) For $p_\phi\geq 0$ one finds, instead, the equilibrium solution 
$\rho=\rho_c=p_\phi/A_\phi(\rho_c,0)$ which yields $\dot{\phi}=0$. 
This solution describes a particle at rest at the distance $\rho_c$ 
for any value of the azimouth. Nearby orbits on the equatorial 
plane, keeping $p_\phi$ constant, arise by perturbing the radial 
velocity $\dot{\rho}\neq 0$ while keeping $\dot{\phi}=\dot{z}=0$. 
In this case, a similar analysis as above shows that nearby orbits 
describe gyrations around the equilibrium solution, which do not 
encircle the central axis. The gyration frequency is now equal to 
$\omega_c=\kappa$, the minimum and maximum distances from the 
axis are found by the two roots $0<\rho_1<\rho_2$ of 
$H=E=V_{eff}(\rho,0)$, while $\dot{\phi}$ has opposite sign 
in the intervals $\rho_1\leq \rho<\rho_c$ and $\rho_c<\rho\leq\rho_2$. 
The Hamiltonian is still given by (\ref{hamepi}), and it is easy 
to check that the lowest order terms quadratic in $z$ are of order 
$O(\xi z^2)$. Thus, the Hamiltonian is again of the general form 
(\ref{ham2}). 

\begin{figure}[h]
\centering
\includegraphics[scale=0.40]{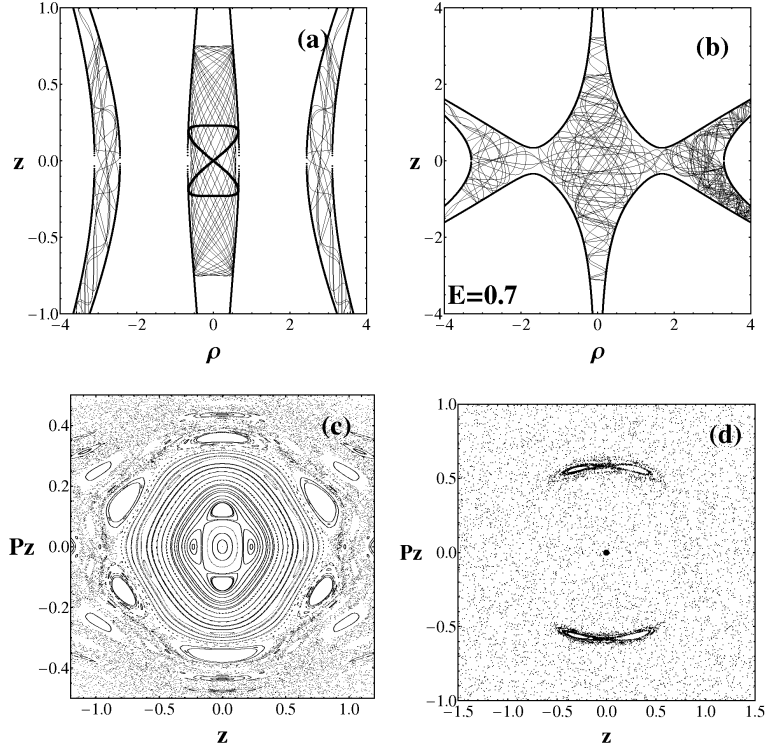}
\caption{\small (a) The main types of orbits for $E=0.2$. The central orbit 
in gray is a typical non-resonant regular orbit undergoing mirror oscillations. 
The `figure 8' orbit (black) is a 2:1 resonant orbit. Both orbits are 
limited by the central curves of zero velocity. On the other hand, the 
left and right gray curves show chaotic orbits limited by the outer 
CZVs. (b) A chaotic orbit for $E=0.7$, exploring the interior domain  
limited by the corresponding CZVs. (c) The surface of section $(z,p_z)$ 
for $\rho=0$, $p_\rho>0$ at the energy $E=0.2$. Non-resonant regular orbits 
belong to invariant curves surrounding the central (equatorial) periodic 
orbit at $z=p_z=0$, while resonant orbits belong to island chains around 
the origin. The domain of stability is surrounded by a sea of chaotic 
orbits. (d) The surface of section at $E=0.7$, where global chaos prevails.  
} \label{orbpc}
\end{figure}
The simplest, albeit without loss of generality, case to consider 
is $p_\phi=0$, which we hereafter focus on. In this case,} the orbital 
motion takes place on a meridian plane $(\rho,z)$ rotating with angular
velocity $\dot{\phi}=-A_\phi/\rho$. Gyrating orbits cross the 
axis $\rho=0$. In order to formally avoid discontinuous transitions 
in the value of $\phi$ at each crossing, the cylindrical radius $\rho$ 
can be adopted to take both positive and negative values. In units in 
which $B_0=2$, $\beta_1=1$ in Eq.(\ref{aphi}), the motion on the meridian 
plane is described by the Hamiltonian:
\begin{equation}\label{hammod}
H(\rho,z,p_\rho,p_z)=\frac{1}{2}(p^2_\rho+p^2_z)+V(\rho,z)
\end{equation}
with a `potential' (equal to $A_{\phi}^2/2$) given by
\begin{equation}\label{pot}
V(\rho,z)=\frac{1}{2}\rho^2+\frac{1}{2}\rho^2z^2
-\frac{1}{8}\rho^4+\frac{1}{8}\rho^2z^4
-\frac{1}{16}\rho^4z^2+\frac{1}{128}\rho^6~~.
\end{equation}

Figure \ref{isopot} plots the equipotential curves, or curves of
zero velocity (CZV), for different values of the energy
$E=H(\rho,z)=$ $V(\rho,z)$ with $p_{\rho}$ = $p_z$ = 0 
(the magnetic field lines have a similar form (see 
figure 1 in Engel et al. (1995)), but form a small angle with the 
lines of Fig.\ref{isopot}, which correspond to constant values of 
$A_\phi^2$ ). 
These curves define the limits that can be reached by the orbits.
The potential along the $\rho$-axis ($z$=0) is $V(\rho,0)=\frac{1}{2}
\rho^2(1-\frac{\rho^2}{8})^2=E$ and has two minima and one maximum 
(in each of the half-planes $\rho>0$ or $\rho<0$). The maximum
is equal to $V(\rho,0)$=$E_{crit}$=16/27$\approx0.5926$ for $\rho$=
$\rho_{crit}$=$\pm\sqrt{8/3}\approx1.633$. 

Examples of orbits in the above system are given in Fig.\ref{orbpc}. 
When $0<E<E_{crit}$ there are three permissible regions of orbits, 
one close to the z-axis ($\rho=0$) and two on the right and on the 
left (Fig.\ref{orbpc}a). In the central region we find ordered orbits, 
which obey some quasi-integral of motion. The gray and black central 
orbits in Fig.\ref{orbpc}a correspond to a non-resonant and resonant 
case respectively. The associated phase portrait (surface of section 
$\rho=0$, $\dot{\rho}>0$) is shown in Fig\ref{orbpc}c, for the energy 
$E=0.2$. In this case, the main resonances are 2:1 and 3:1, which both 
produce double chains of islands of stability (four and six respectively 
in Fig.\ref{orbpc}c), as discussed in detail in section 4. On the 
other hand, in Fig.\ref{orbpc}a the left and right orbits in gray 
are chaotic orbits having no intersection with the surface of 
section of Fig.\ref{orbpc}c. Finally, for much higher energies 
($E=0.7$, Figs\ref{orbpc}b,d), the central orbit has become 
unstable and the system is characterized by global chaos. 

In section 4, we shall employ the normal form method in order to 
compute the critical energy value where the onset of global chaos 
takes place. 

\section{Non-resonant normal form}

\subsection{Algorithm}
The Hamiltonian (\ref{hammod}) is of the general form (\ref{ham1}).
A non-resonant normal form for this Hamiltonian can be constructed 
by the method of Dragt and Finn (1979) or Engel et al. (1995). 
We summarize here the main steps, introducing our own notation and 
terminology:\\
\\
i) {\it Introduction of complex canonical variables.} Introducing
the linear canonical change of variables
\begin{equation}\label{trans}
\rho={q_1+ip_1\over \sqrt{2\omega_{1,0}}},~~~p_{\rho}={iq_1+p_1\over
\sqrt{2/\omega_{1,0}}},\\ z=q_2,~~~p_z=p_2
\end{equation}
with $\omega_{1,0}$ equal to the frequency induced by the quadratic
term $O(\rho^2)$ of the potential ($\omega_{1,0}=1$ in our example),
the Hamiltonian takes the form $H=H_2+H_4+H_6$, with
\begin{equation}\label{hamquad}
H_2(q_1,p_1,p_2)=i\omega_{1,0} q_1p_1+\frac{1}{2}p^2_2~~.
\end{equation}
The terms $H_4$ and $H_6$ are of fourth and sixth degree respectively.
For orbits crossing the z-axis (with $p_\phi=0$), $\omega_{1,0}$ is
equal to {\it half} the gyration frequency.\\
\\
ii) {\it Book-keeping:} We organize the terms in the Hamiltonian
in groups of `different order of smallness'. Formally, we introduce
a `book-keeping' parameter $\lambda$, with numerical value
$\lambda=1$, and write the Hamiltonian as
\begin{equation}\label{hambk}
H\equiv H^{(0)}=H_0^{(0)} +\lambda H_1^{(0)}+\lambda^2 H_2^{(0)}+...
\end{equation}
The superscript $(0)$ means `no normalization step performed so far'.
A subscript $i$, accompanied by a book-keeping coefficient
$\lambda^i$, means ``i-th order of smallness''.  The arrangement of
all the Hamiltonian terms in the groups $H^{(0)}_0$, $H^{(0)}_1$,
etc., is done by adopting a `book-keeping rule'. For polynomial
Hamiltonian models, the simplest choice of rule is to associate
book-keeping order with polynomial degree. Thus, in the Hamiltonian
(\ref{hambk}) we set $H^{(0)}_s$ to be the ensemble of terms of
polynomial degree $2s+2$. This renders our non-resonant construction
equivalent to those of Dragt and Finn (1979) or Engel (1995).
However, the above book-keeping rule will be modified in the resonant
construction dealt with in section 4 below. We note here that the 
form of the Hamiltonian (section 2) in the more general case 
$p_\phi>0$ allows for employing the same book-keeping rule as 
above, with $\rho$ substituted by the local variable $\xi=\rho-\rho_c$ 
around an equatorial orbit at $\rho=\rho_c$. Instead, in the 
case $p_\phi<0$ we have to add one more power of $\lambda$ 
for every power of the quantity $\beta_1$, which now plays also 
the role of small parameter.\\
\\
iii) {\it Choice of `kernel set' ${\cal M}$.} To choose a `kernel set' 
means to answer the question of which terms are kept in the normal form 
along the normalization process. In the present example, in the
Hamiltonian there appear monomial terms of the form
$q_1^{k_1}p_1^{l_1}q_2^{k_2}p_2^{l_2}$, where $k_1,~l_1,~k_2,~l_2\geq0$. 
For determining a non-resonant formal integral, it suffices to keep 
in the normal form the terms satisfying $k_1=l_1$. 
Then, the normal form takes the form $Z\equiv
Z(q_1p_1,q_2,p_2)$, i.e. its dependence on $q_1,p_1$ is only through
the product $q_1p_1$. Then, the quantity $I_1=iq_1p_1$ is a formal
integral, since $\{I_1,Z\}=0$ (where $\{\cdot,\cdot\}$ denotes the
Poisson bracket).  However, it is practical to exclude some further
terms from the  normal form. Namely, we also exclude the terms with
$k_1=l_1$ but $l_2\neq 0$, except if $k_1=l_1=0$ and $l_2=2$,
$k_2=0$. This means to retain terms of the form $I_1^n q_2^{k_2}$,
for exponents $n\neq 0$, as well as the `kinetic' term
$\frac{1}{2}p_2^2$. Since $I_1$ is a formal integral, after these
definitions the normal form reduces to the form $Z=\omega_1 I_1 +
p_2^2/2 + U(I_1,q_2)$, i.e. $Z$ takes the form of an one-degree of
freedom oscillator ($\omega_1 I_1$) plus a `kinetic' and a
`potential' term for the second degree of freedom (for which $I_1$
acts as a parameter in the `potential' $U(I_1,q_2)$).

In summary, as kernel set ${\cal M}$ we choose:
\begin{equation}\label{resmod}
 {\cal M} = \left\{k_1=l_1~\mbox{and}~
\left(k_2=0,l_2=2~\mbox{if}~k_1+l_1=0,~
\mbox{or}~l_2=0~\mbox{if}~k_1+l_1>0\right)\right\}~~.
\end{equation}

iv) {\it Normalization.}
We construct the non-resonant normal form by using canonical
transformations via Lie series (see, e.g., Efthymiopoulos 2012 for a
tutorial introduction to Lie series). We thus introduce the sequence
of transformations: $(q_1,q_2,p_1,p_2)$ $\rightarrow$
$(q_1^{(1)},q_2^{(1)},p_1^{(1)}, p_2^{(1)})$, $\rightarrow$
$(q_1^{(2)},q_2^{(2)},p_1^{(2)}, p_2^{(2)})$, $\rightarrow$,
$\ldots$, where superscripts $(1)$, $(2)$,...,denote the new canonical
variables after the first, second, etc., normalization steps. In the
$r$-th step, the transformation is given in terms of a Lie generating
function $\chi_r(q_1^{(r)},q_2^{(r)},p_1^{(r)}, p_2^{(r)})$. The
transformation of any function $f(q_1^{(r-1)},q_2^{(r-1)},
p_1^{(r-1)}, p_2^{(r-1)})$ in the new variables is found by
replacing $(q_1^{(r-1)},q_2^{(r-1)},p_1^{(r-1)}, p_2^{(r-1)})$
with $(q_1^{(r)},q_2^{(r)},p_1^{(r)}, p_2^{(r)})$ in the arguments
of $f$ and, then, by computing $f'=\exp(L_{\chi_r})f$, where
\begin{equation}\label{expoper}
\exp(L_{\chi_r}) = \sum_{k=0}^{\infty}{1\over k!}L^k_{\chi_r}~~,
\end{equation}
$L_{\chi_r}=\{\cdot,\chi_r\}$ denoting the Poisson bracket operator.
In particular, the variables themselves are transformed according
to $q_1^{(r)}=\exp(L_{\chi_r})q_1^{(r-1)}(q^{(r)})$, where
$q_1^{(r-1)}(q^{(r)})$ is the identity function (and similarly for the
remaining variables). In the sequel, for simplicity we drop superscripts
in the notation of the canonical variables.

The generating functions $\chi_r$, $r=1,2,\ldots$ are computed recursively.
Namely, the Hamiltonian after $r-1$ normalization steps has the form
\begin{equation}\label{hrm1}
H^{(r-1)}=Z_0 + \lambda Z_1 + \lambda^2 Z_2 +\ldots +
\lambda^{r-1}Z_{r-1} +\lambda^{r}H^{(r-1)}_r
+\lambda^{r+1}H^{(r-1)}_{r+1}+\ldots
\end{equation}
where all the terms in $Z_0$, $Z_1$,...,$Z_{r-1}$ are in normal
form, i.e., they belong to ${\cal M}$. According to the book-keeping
rule choosen in (ii), the terms $Z_s$, or $H^{(r-1)}_s$ are of
polynomial degree $2s+2$. In particular, $Z_0\equiv H_2=i\omega_{1,0}q_1p_1
+ p_2^2/2$. Let $\tilde{H}^{(r-1)}_r$ denote the terms of $H^{(r-1)}_r$
not belonging to ${\cal M}$. The generating function $\chi_r$
is the solution to the homological equation
\begin{equation}\label{homo}
\{Z_0,\chi_r\}=\left\{i\omega_{1,0}q_1p_1
+ {p_2^2\over 2},\chi_r\right\}= -\lambda^r\tilde{H}^{(r-1)}_r~~.
\end{equation}
After computing $\chi_r$, we compute the transformed Hamiltonian
\begin{equation}\label{hrlie}
H^{(r)}=\exp(L_{\chi_r})H^{(r-1)}~~.
\end{equation}
This is in normal form up to terms of book-keeping order $r$, namely
\begin{equation}\label{hr}
H^{(r)}=Z_0 + \lambda Z_1 + \lambda^2 Z_2 +\ldots +
\lambda^{r}Z_{r} +\lambda^{r+1}H^{(r)}_{r+1}
+\lambda^{r+2}H^{(r)}_{r+2}+\ldots
\end{equation}
This completes one step of the normalization algorithm.

In order to solve the homological equation (\ref{homo}), we first
decompose $\tilde{H}^{(r-1)}_r$ in the sum of monomials
$$
\tilde{H}^{(r-1)}_r=\sum_{\begin{array}{c}^{k_1,l_1,k_2,l_2>0}\\
^{k_1+l_1+k_2+l_2=2r+2}\end{array}}
h_{k_1,l_1,k_2,l_2}^{(r-1)}q_1^{k_1}p_1^{l_1}q_2^{k_2}p_2^{l_2}
$$
with known coefficients $h_{k_1,l_1,k_2,l_2}^{(r-1)}$. However,
one notes that, contrary to the usual Birkhoff normal form around
elliptic equilibria, in the present case, due to the presence of
the term $p_2^2/2$ in $Z_0$, a monomial
$q_1^{k_1}p_1^{l_1}q_2^{k_2}p_2^{l_2}$ does not constitute an
eigenfunction of the linear operator $D_{Z_0}=\{Z_0,\cdot\}$,
since
\begin{eqnarray} \label{domeiga}
D_{Z_0}q_1^{k_1}p_1^{l_1}q_2^{k_2}p_2^{l_2} =
i(l_1-k_1)\omega_{1,0}q_1^{k_1}p_1^{l_1}q_2^{k_2}p_2^{l_2}
-k_2q_1^{k_1}p_1^{l_1}q_2^{k_2-1}p_2^{l_2+1}~~~.
\end{eqnarray}
Thus, the homological equation cannot be solved by term-by-term
comparison of the coefficients as in the usual Birkhoff case.
Instead, we form the sets of terms
\begin{eqnarray}\label{block}
{\cal A}_{kl}&=&\Bigg\{a^{(r-1)}_{kl,n}q_1^kp_1^lq_2^np_2^{2r+2-k-l-n}:
k+l\leq 2r+2,\\ \nonumber
&~&~~~~~~~~~~~~n=0,1,\ldots, 2r+2-k-l\Bigg\}
\end{eqnarray}
where $a^{(r-1)}_{kl,n}\equiv-h^{(r-1)}_{k,l,n,2r+2-k-l-n}$ for
given values of $k=k_1$ and $l=l_1$. Then, setting the generating
function $\chi_r$ to contain a similar group of terms with (yet
unknown) coefficients $b^{(r-1)}_{kl,n}$, the homological equation
is decomposed in the set of linear systems of equations
\begin{equation}\label{bidiag}
\left(
\begin{array}{ccccl}
c_{kl} &   1     &    ~  &    ~   &  ~      \\
   ~   & c_{kl}  &   2   &    ~   &  ~      \\
   ~   &   ~     &\ddots & \ddots &  ~      \\
   ~   &   ~     &   ~   &c_{kl}  & 2r+2-k-l\\
   ~   &   ~     &   ~   &   ~    & c_{kl}  \\
\end{array}
\right) \left(
\begin{array}{c}
b^{(r-1)}_{kl,0}\\
b^{(r-1)}_{kl,1}\\
\vdots\\
\vdots\\
b^{(r-1)}_{kl,2r+2-k-l}\\
\end{array}
\right)
=\left(
\begin{array}{c}
a^{(r-1)}_{kl,0}\\
a^{(r-1)}_{kl,1}\\
\vdots\\
\vdots\\
a^{(r-1)}_{kl,2r+2-k-l}\\
\end{array}
\right)
\end{equation}
where $c_{kl}=i(l-k)\omega_{1,0}$. If $l\neq k$, the system
(\ref{bidiag}) can be solved by backward substitution, i.e. we first
solve the last equation, then substitute and solve the previous one,
etc. On the other hand, the choice of kernel set ${\cal M}$ implies that
the normalization should eliminate from the normal form also terms
with $k=l$ and $n=0,1,\ldots 2r+1-k-l$. Then we have
$a^{(r-1)}_{kl,2r+2-k-l}=0$ and the last of Eqs.(\ref{bidiag})
becomes the identity $0=0$, while the remaining equations for $k=l$
form a diagonal system with non zero-determinant for the
coefficients $b^{(r-1)}_{kl,n}$ with $n=1,2,\ldots 2r+2-k-l$ in
terms of the coefficients $a^{(r-1)}_{kl,n}$ with $n=0,1,\ldots
2r+1-k-l$. Also, the coefficient $b^{(r-1)}_{kl,0}$ becomes
arbitrary, and can be set equal to zero. This completely specifies
the generating function $\chi_r$.

\begin{figure}[h]
\centering
\includegraphics[scale=0.35]{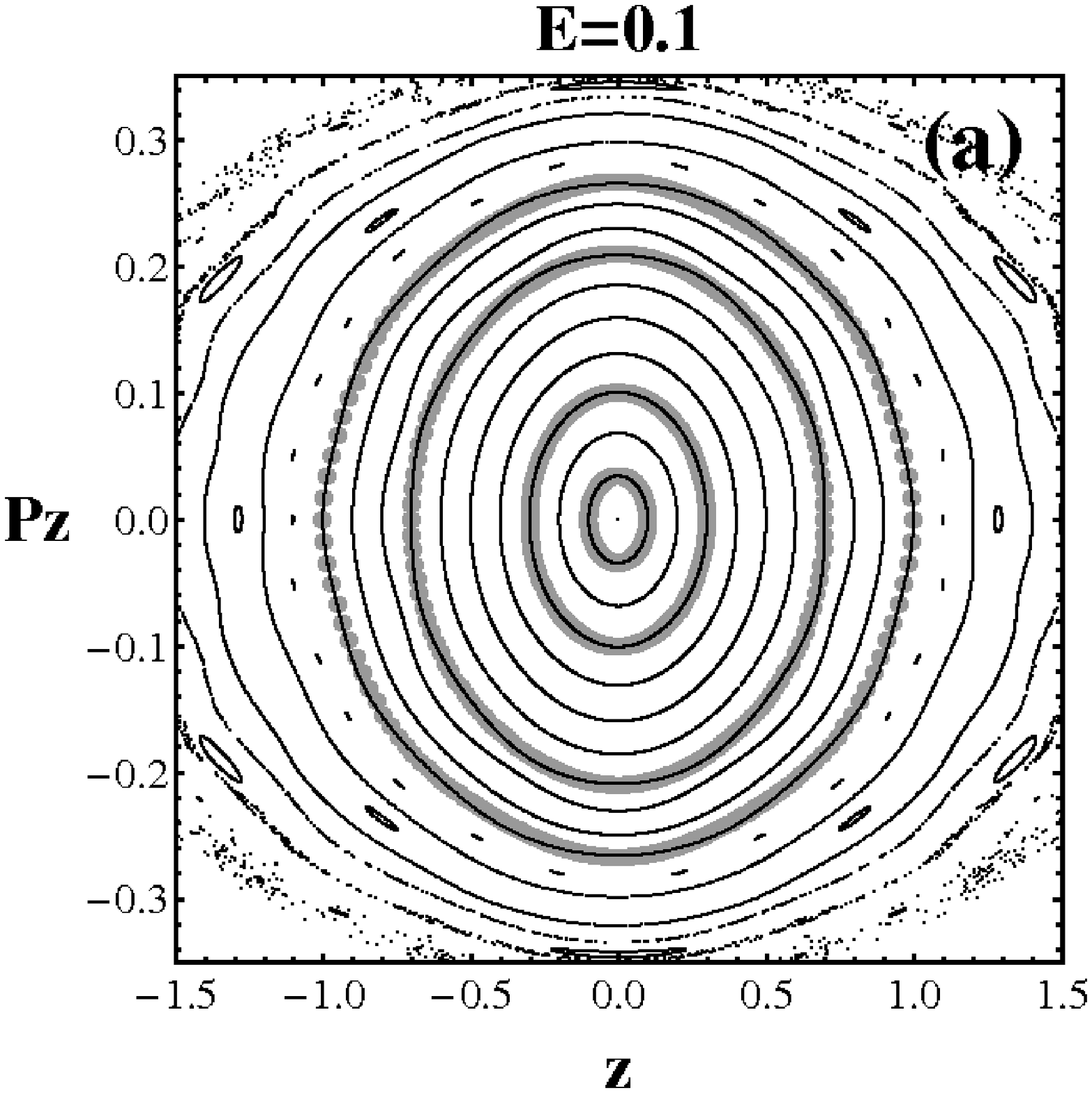}
\includegraphics[scale=0.35]{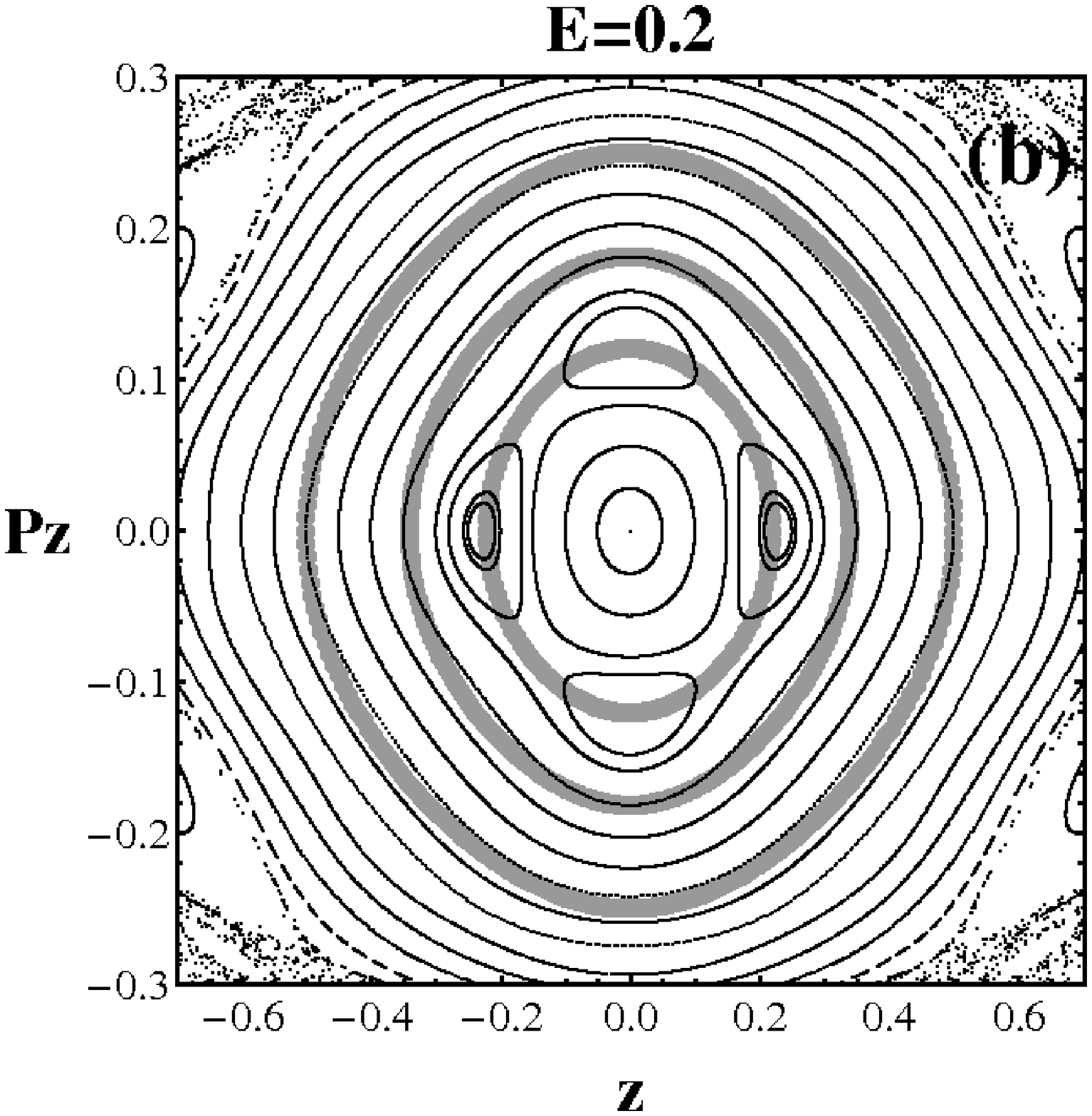}
\caption{\small (a) The Poincar\'{e} surface of section for $E=0.1$ (black
curves) superimposed with the curves of equal values of the
non-resonant formal integral (gray points), truncated at order 5
(corresponding to polynomial degree 12). (b) Same as in (a) but for
$E=0.2$. The non-resonant formal integral cannot reproduce the
2:1 islands of stability.} \label{nonresi}
\end{figure}
\subsection{Implementation}
Implementing the above procedure to the Hamiltonian (\ref{hambk}), we
compute a non-resonant normal form using a computer-algebraic program.
At low normalization orders, this yields results similar to those of
Engel et al. (1995). However, we extended the computations up to the
normalization order $r=15$ (corresponding to the polynomial degree 32),
while all series expressions were truncated at order $r_{trunc}=20$
(corresponding to polynomial degree 42). The asymptotic behavior of the
series at high normalization orders is examined in the next subsection.
At any rate, even at low normalization orders one obtains expressions
for the non-resonant formal integral comparing well with numerical
results for the non-resonant ordered orbits. As an example, after five
normalization steps, the transformed Hamiltonian (restoring the book-keeping
constant value $\lambda=1$) reads:
\begin{eqnarray}\label{hamnorm}
H^{(5)}&=&I_1 - 0.1875 I_1^2 - 0.046875 I_1^3 - 0.0256348 I_1^4
- 0.0184021 I_1^5 - 0.0152607 I_1^6 ~~~~~~~~~~~~~\nonumber\\
&+& 0.5 p_2^2 + 0.5 I_1 q_2^2 +
0.15625 I_1^2 q_2^2 + 0.1875 I_1^3 q_2^2
+ 0.299194 I_1^4 q_2^2 + 0.551285 I_1^5 q_2^2 \nonumber\\
&-&0.151042 I_1^2 q_2^4 - 0.46224 I_1^3 q_2^4
-1.29767 I_1^4 q_2^4 \\
&+&0.107812 I_1^2 q_2^6 + 0.669227 I_1^3 q_2^6 - 0.0697545 I_1^2 q_2^8
+R^{(5)}\nonumber
\end{eqnarray}
where $I_1=iq_1p_1$, and $R^{(5)}=H^{(5)}_6+H^{(5)}_7+...$ denotes
the remainder series. The quantity $I_1$ represents a formal
integral of motion, i.e., the non-resonant formal integral. The
mirror frequency is expressed in terms of $I_1$ by the terms in
(\ref{hamnorm}) quadratic in $q_2$. Thus
\begin{equation}\label{ome2i}
\omega_2^2(I_1)=I_1 + 0.3125 I_1^2 + 0.375 I_1^3 + 0.598388I_1^4
+ 1.10257 I_1^5+...
\end{equation}
As explained below, the possibility to compute $\omega_2$ in terms of $I_1$
turns to be crucial in the subsequent construction of a resonant normal form 
(section 4).

In the expression (\ref{hamnorm}) all symbols refer to the new canonical
variables after the composition of five canonical transformations, i.e.
$q_1\equiv q_1^{(5)}$ etc. Similarly, $I_1$ expresses the quasi-integral 
in the new canonical variables. This can be transformed to the original 
variables by computing (up to book-keeping order r) the expression
\begin{equation}\label{phiold}
\Phi(q_1,p_1,q_2,p_2)=
\exp(-L_{\chi_1})\circ\exp(-L_{\chi_2})\ldots\circ\exp(-L_{\chi_r})(iq_1p_1)
~~~.
\end{equation}
Finally, the integral $\Phi$ can be expressed in the original variables
$\rho,p_\rho,z,p_z$ by the substitution
\begin{equation}\label{orvar}
q_1=\frac{1}{2}\sqrt{2}(\rho-i p_{\rho}), ~~
p_1=\frac{1}{2} \sqrt{2}(p_{\rho}-i \rho),~~ \\
q_2=z,~~~p_2=p_z~~.
\end{equation}
This allows to compute theoretical invariant curves on a surface of
section $\rho=0$, $p_\rho>0$ for given energy $E$. To this end, we
set $\rho=0$, $p_{\rho}=(2(E-V(0,z))-p^2_z)^{1/2}$. Then, $\Phi$ is
expressed as a function of $(z,p_z)$ only, i.e.
$\Phi=\Phi_{sect}(z,p_z;E)$. Figure \ref{nonresi}a shows a
comparison between theoretical and numerical invariant curves on the
surface of section for $E=0.1$ (gray points and black curves
respectively). The theoretical curves are obtained by computing the
level curves $\Phi_{sect}(z,p_z;E)=I_{ct}$, where the constant value
$I_{ct}$ is computed as $I_{ct}= \Phi_{sect}(z_0,p_{z0};E)$,
$(z_0,p_{z0})$ being the initial conditions leading to one invariant
curve. In Fig.\ref{nonresi}a the normalization order is relatively
low ($r=5$). Even so, the theoretical invariant curves have a good
degree of coincidence with the numerical curves in nearly the whole
stability domain. In fact, at this energy level no conspicuous
resonances are present in the interior of the stability domain,
while important resonances are only present near its boundary.  The
situation, however, is altered at higher energies, as exemplified in
Fig.\ref{nonresi}b, referring to the surface of section for $E=0.2$.
In this case, the numerical surface of section exhibits two couples
of conspicuous islands corresponding to a (double) 2:1 resonance. As
shown in section 4, the value of the critical energy $E_{2:1}$ where
the 2:1 periodic orbits bifurcate from the center (at $z=p_z=0$), as
well as the existence of a double 2:1 island chain, can be predicted
already by the non-resonant normal form. We find that the energy 
$E=0.2$ is a little above the bifurcation value and the 2:1 island 
chains have moved from the center outwards. However, it is obvious 
that the non-resonant normal form construction is unable to reproduce 
the phase portrait in the zone of the 2:1 resonance. Instead, the 
non-resonant theoretical invariant curves pass through the resonant 
islands, i.e. they mimic a non-resonant behavior. This problem is 
remedied in section 4 by the construction of a resonant normal form.

\subsection{Asymptotic behavior}
The (non) convergence behavior of formal series, in general, is
determined by the accumulation, in the series terms at successive
orders, of {\it small divisors}. In the usual Birkhoff series, the
pattern of accumulation of divisors can be unravelled by carefully
examining the various chains of terms produced by the recursive
normalization scheme at successive orders (see, for example,
Efthymiopoulos et al. 2004 for a heuristic analysis of the
accumulation of divisors in both cases of a non-resonant and
resonant Birkhoff normal form around an elliptic equilibrium). In
the present case of non-resonant normal form, however, the
analysis is perplexed by the fact that the propagation of divisors
depends on the solutions, at each order $r$, of the non-diagonal set
of Eqs.(\ref{bidiag}), due to the fact that the original Hamiltonian
does not contain a term $i\omega_2p_2q_2$. Still, one readily sees
that implementing repeatedly Eqs.(\ref{bidiag}) at successive orders
leads to chains of terms growing in size with $r$ by an upper bound
$Q_r(\epsilon)=O(r!^a(\epsilon^b/\omega_{1,0}^c)^r)$, where $a,b,c$
are positive exponents and $\epsilon=(|p_1q_1|+|p_2^2+q_2^2|)^{1/2}$
is a measure of the distance, in phase space, from the origin. This
implies that an asymptotic behavior is expected for the adiabatic
invariant formal series. Hereafter we demonstrate that this is so by
numerical experiments where, for fixed value of the energy $E$, we
study the behavior of the series at high orders as a function of the
energy $\Delta E$ of the mirror oscillation, given by $\Delta
E=E-E_1$, where $E_1$ is the gyration kinetic energy. The energy
$\Delta E$ plays here the role of a small parameter, since for
$\Delta E=0$ we have equatorial orbits corresponding to a
one-degree-of freedom, i.e. an integrable limit of the Hamiltonian
model (\ref{hammod}).

\begin{figure}
\centering
\includegraphics[scale=0.55]{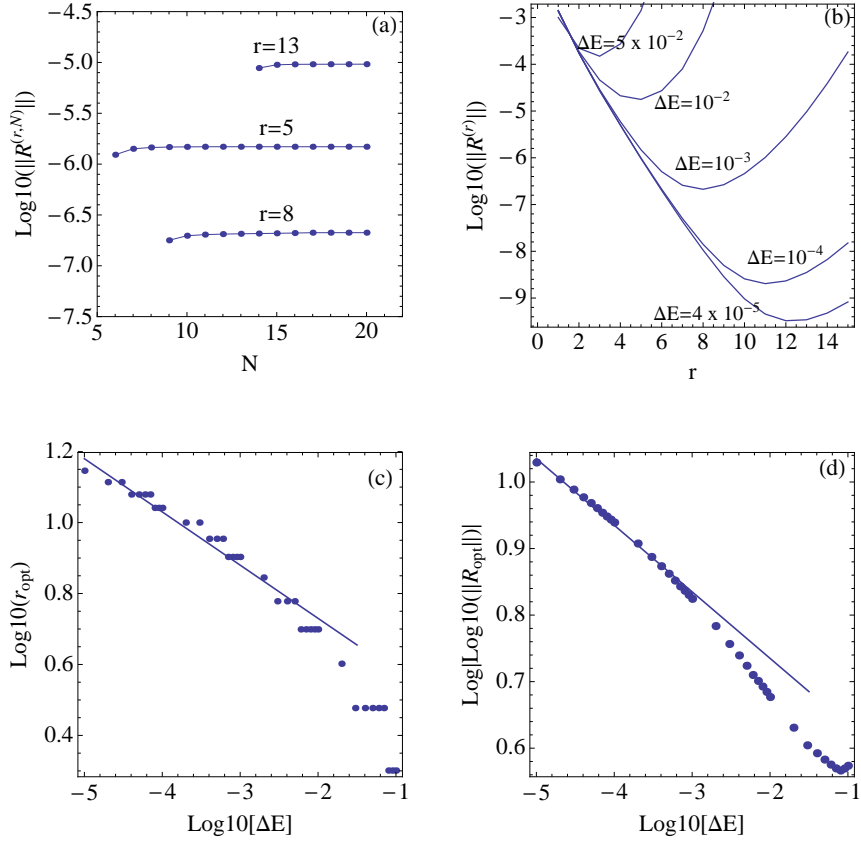}
\caption{\small Asymptotic behavior of the non-resonant normal
form series exemplified by a specific computation using the norm
definition (\ref{rnorm}) for the remainder function with parameters
$\beta=0$, and $E=0.2$. (a) Dependence of the truncated remainder
norm on the truncation order $N$ at three different normalization
orders $r=5$, $r=8$, and $r=13$, when $\Delta E=0.001$. (b) The
quantity $||R^{(r,20)}||$ as a function of $r$ for the indicated
values of $\Delta E$. The optimal remainder corresponds to the
minimum of each curve. (c) The optimal order as a function of
$\Delta E$ in log-log scale. An approximate inverse power-law holds
for $\Delta E$ below a threshold $\Delta E\approx10^{-3}$. The
fitting curve corresponds to an exponent 0.15. (d) The value of the
optimal remainder (in $\log|\log|$ scale) versus $\Delta E$ (in log
scale). The straight line represents an exponential law
$R_{opt}\propto \exp((-\Delta E_0/\Delta E)^d)$ with $\Delta
E_0=10^{-3}$ and $d=0.12$.} \label{nrasym}
\end{figure}
To this end, we consider first a norm definition for the remainder series.
At the normalization order $r$, the remainder series has the form (setting
$\lambda=1$):
\begin{equation}\label{remseries}
R^{(r)}(q_1,q_2,p_1,p_2)=\sum_{s=r+1}^{\infty}H^{(r)}_s =
\sum _{s=r+1}^{\infty}
\sum_{\begin{array}{c}^{k_1,l_1,k_2,l_2>0}\\
^{k_1+l_1+k_2+l_2=2s+2}\end{array}}
G_{s,k_1,l_1,k_2,l_2}^{(r)}q_1^{k_1}p_1^{l_1}q_2^{k_2}p_2^{l_2}~~.
\end{equation}
We define the N-th order truncation of $R^{(r)}$ as
\begin{equation}\label{remseries}
R^{(r,N)}(q_1,q_2,p_1,p_2)=\sum_{s=r+1}^{N}H^{(r)}_s =
\sum _{s=r+1}^{N}
\sum_{\begin{array}{c}^{k_1,l_1,k_2,l_2>0}\\
^{k_1+l_1+k_2+l_2=2s+2}\end{array}}
G_{s,k_1,l_1,k_2,l_2}^{(r)}q_1^{k_1}p_1^{l_1}q_2^{k_2}p_2^{l_2}~~.
\end{equation}
Let now $E,\Delta E$ be fixed such that $0\leq\Delta E<E$. Consider
a fixed direction $q_2=\beta p_2$ in the plane $(q_2,p_2)$. It is
easy to prove that the quantity
\begin{eqnarray}\label{rnorm}
||R^{(r,N)}||_{E,\Delta E,\beta} &=&
\sum _{s=r+1}^{N}
\sum_{\begin{array}{c}^{k_1,l_1,k_2,l_2>0}\\
^{k_1+l_1+k_2+l_2=2s+2}\end{array}}
\Bigg(
|G_{s,k_1,l_1,k_2,l_2}^{(r)}|
((E-\Delta E)/\omega_{1,0}))^{k_1+l_1\over 2}\nonumber\\
&\times&
|\beta|^{k_2}
\left({2\Delta E\over 1+\beta^2\omega_2^2((E-\Delta E)/\omega_{1,0}))}
\right)^{k_2+l_2\over 2}\Bigg)
\end{eqnarray}
satisfies all the properties of norm definition. The norm (\ref{rnorm})
provides a measure of the size of the remainder at the given energy levels
$E,\Delta E$. In particular, we find the following:

i) For $\Delta E$ sufficiently small, the sequence
$||R^{(r,N)}||_{E,\Delta E,\beta}$, for $N=r+1$, $r+2$, $\ldots$ is
convergent for, $N\rightarrow\infty$, at all normalization orders
$r$. An example is given in Fig.\ref{nrasym}a. We fix $\beta=0$,
$E=0.2$, $\Delta E=0.001$, referring to an estimate of the size of
the remainder along the axis $z=0$ on the surface of section of
Fig.\ref{nonresi}a, at a distance $p_z=\sqrt{2\Delta E}\approx
4.5\times 10^{-2}$. The behavior of $||R^{(r,N)}||$ is shown for
three different normalization orders $r=5$, $r=8$, $r=13$. In all
three cases, we find that the norm of the truncated remainder
converges rather quickly with the truncation order $N$. This implies
that the value of the remainder found at the maximum truncation
order $N=20$ used here is a good measure of the limiting value
$||R^{(r,\infty)}||$ for all three chosen normalization orders $r$.

ii) Figure \ref{nrasym}a provides an indication of the asymptotic behavior
of the series for the particular parameters. Namely, we observe that the
norm of the remainder decreases as we move from the normalization order
$r=5$ to $r=8$, however, it increases as we move from $r=8$ to $r=13$.
The dependence of $||R^{(r,N)}||$ on $r$ (fixing $N$ to the maximum $N=20$),
for fixed $\beta,E$ is shown in detail in Fig.\ref{nrasym}b, for five
different values of the mirror oscillation energy $\Delta E$.
The asymptotic behavior of the remainder series is evident in this plot,
which shows also that the optimal order $r_{opt}$, at which the norm of
the remainder becomes minimum, decreases as $\Delta E$ increases,
while the value of the norm at the optimal order increases with
$\Delta E$.

iii) The dependence of $r_{opt}$ on $\Delta E$, shown in
Fig.\ref{nrasym}c is approximately power-law like. The straight line
indicates a power law with exponent -0.15 which holds for energies
below a threshold value $\Delta E\approx10^{-3}$. Note that an
optimal order as low as $r_{opt}=2$ is reached at the energy $\Delta
E\approx 6\times 10^{-2}$. This corresponds to $p_z\approx \pm 0.35$
along the axis $z=0$ on the surface of section. Simple visual
inspection of Fig.\ref{orbpc}c shows that this value is close to
the border of the stability domain for $z=0$. This suggests that the
limits of the stability domain can be estimated analytically by the
requirement that $r_{opt}$ becomes very small., e.g. $r_{opt}=2$.
Physically, this marks the limit of overall validity of the
non-resonant normal form.

iv) The optimal remainder value is exponentially small in $1/\Delta
E$. Figure \ref{nrasym}d shows the quantity
$\log|\log(||R^{(r_{opt},20)}||)|$ vs. $\log\Delta E$ (the value of
the remainder decreases for higher values in the ordinate). The
straight line corresponds to a law
$||R^{(r_{opt})}||\approx\exp(-(\Delta E_0/\Delta E)^d)$ with
$d=0.12$. Notice that, as shown also in Fig.\ref{nrasym}b, the
exponential dependence implies that for $\Delta E$ small we can
obtain quite small optimal remainder values (e.g. of order $10^{-9}$
when $\Delta E=10^{-4}$, rising to $10^{-4}$ when $\Delta E$ is of
order $10^{-2}$. These numbers set the overall level of precision of
the non-resonant quasi-integrals as a function of the mirror
oscillation energy.

Let us note here that the asymptotic character of the above 
normal form construction is due to the fact that we seek an expansion 
in which all quantities are defined in an open domain of the phase space. 
If, instead, after a few normalization steps, we fix a value of the 
action $I_1=I_1*$, and expand locally the Hamiltonian (e.g. Eq.(\ref{hamnorm})) 
around this value, we can arrive at a form of the normalized Hamiltonian 
in action-angle variables, allowing for the implementation of the 
Kolmogorov algorithm of the KAM theorem. The existence of invariant 
curves in the phase portraits indicates that such a process should 
yield convergent series on a Cantor set of initial conditions. However, 
exploring such convergence is beyond the scope of our present study.

\section{Resonant normal form}
\subsection{Bifurcation energy}
Resonant periodic orbits $m_2/m_1$ bifurcate from the central
equatorial orbit when $m_2/m_1=\omega_2/\omega_1$. Using the
non-resonant normal form we can predict the bifurcation energy of
the $m_2$:$m_1$ family. The gyro-frequency of the equatorial orbit
$\omega_{1,eq}$ is computed by setting $q_2=p_2=0$ in the normal
form (as in Eq.(\ref{hamnorm})). Then
\begin{equation}\label{ome1eq}
\omega_{1,eq}(I_1)=\frac{\partial Z(I_1,q_2=p_2=0)}{\partial I_1}~~.
\end{equation}
The $m_2/m_1$ family bifurcates at the action value $I_{1}^*$ given
by:
\begin{equation}\label{bifact}
m_2\omega_{1,eq}(I_1^*)=m_1\omega_2(I_1^*)~~
\end{equation}
where $\omega_2(I_1)$ is computed as in Eq.(\ref{ome2i}). Finally,
the bifurcation energy is computed by $E_{m2/m1}=Z(I_1^*)$.

\subsection{Algorithm}
Denoting $\omega_1^*=\omega_1(I_1^*)$, $\omega_2^*=\omega_2(I_1^*)$,
we now construct a resonant normal form for the $m_2/m_1$ resonance
as follows:\\

i) {\it Book-keeping and Hamiltonian preparation:} As long as $I_1$
is considered as a small quantity, one has that the quantities
$\omega_{1,0}-\omega_1=O(I_1)$, $\omega_2^2=O(I_1)$ are also small.
Both quantities play the role of `detuning' parameters (Pucacco et
al. 2008), since they both represent a difference from the
unperturbed frequencies which are $\omega_{1,0}$ and zero
respectively. Fixing a value $I_1=I_1^*$, we can formally take this
fact into account in the Hamiltonian by {\it adding and substracting
the above quantities with a different book-keeping factor}, i.e. we
set:
\begin{eqnarray}\label{hamtrick}
H=i\omega_{1,0}p_1q_1+{1\over 2}p_z^2+...= i\omega_1^*p_1q_1+{1\over
2}p_z^2 + {1\over 2}(\omega_2^*)^2z^2 \nonumber\\
-\lambda\left(i(\omega_1^*-\omega_{1,0})q_1p_1 +{1\over
2}(\omega_2^*)^2z^2+H_4\right) + O(\lambda^2)
\end{eqnarray}

Since $\lambda=1$, nothing has really changed with respect to the
original Hamiltonian. However, the second frequency was now
explicitly introduced in the zero order Hamiltonian term which is
subsequently used in the normal form construction. Furthermore, if
$\omega_1^*$ and $\omega_2^*$ satisfy a resonant condition, it is
possible to proceed with the resonant form of the Birkhoff normal
form. We note that the correspondence between book-keeping orders
and polynomial degrees is now broken, namely, at the book-keeping
order $r$ one has terms of the polynomial degrees $2,4,\ldots,2r+2$.
However, this poses no formal obstacles to the construction of the
normal form.
\footnote{The following comment offers some insight into the 
whole above `book-keeping' process: in the case of the 2:1 resonance 
treated in detail below, we find $|(\omega_1^*-\omega_{1,0})| 
\approx 0.1$, $(\omega_2^*)^2\approx 0.25$. The first quantity 
can be called ``of order 1'', but, for the second, the characterization 
``order 0'' or ``order 1'' would be equally acceptable in practice. 
Note, however, that the fourth order terms in the Hamiltonian include 
a term $h_{22}\rho^2z^2$, with a real coefficient $h_{22}$ and 
$\rho\sim(2I_1)^{1/2}$. One then finds that the value of $h_{22}$ 
has to be such that at the particular radius $\rho_*$, corresponding 
to the resonant value $I_1^*$, one will obtain that $h_{22}(\rho^*)^2$ 
has only a small difference (`of order one') from $2(\omega_2^*)^2)$. 
This is reflected by our choice of book-keeping, which results in the 
quantity $\lambda [h_{22}\rho^2-(1/2)(\omega_2^*)^2]z^2$ formally 
appearing in the Hamiltonian (\ref{hamtrick}). }

The remaining steps in the normal form construction are standard 
(see Efthymiopoulos 2012 for a review). Introducing the linear canonical 
change of variables
\begin{equation}
 z=\frac{ q_2 + ip_2}{\sqrt{2\omega^*_2}}~~~,~~~
     p_z= \frac{\sqrt{\omega^*_2}(iq_2 + p_2)}{ \sqrt{2}}
\end{equation}
the Hamiltonian becomes of the general form (\ref{hambk}) with
\begin{equation}\label{hamquad}
H_0^{(0)}=i\omega^*_1 q_1p_1+i\omega^*_2 q_2p_2~~.
\end{equation}
The functions $H^{(0)}_r$ contain monomials of the form
$q_1^{k_1}p_1^{l_1}q_2^{k_2}p_2^{l_2}$. The book-keeping rule is
$2\leq k_1+k_2+ l_1+l_2 \leq 2r+2$.

ii) {\it Choice of kernel set {\cal M}.} For the resonance
$m_2/m_1$ we set
\begin{equation}\label{resmod1}
{\cal M}_{res} = \left\{(k_1,k_2,l_1,l_2)~\mbox{such that~}
(k_1 - l_1)m_1 + (k_2 - l_2)m_2 = 0\right\}~~.
\end{equation}
This choice ensures that the quantity
$I_{res}=i(m_1q_1p_1+m_2q_2p_2)$ is a formal integral in the new
canonical variables.

iii) {\it Normalization}. The normalization proceeds recursively by
the same scheme as in subsection 4.1. The functions
$\tilde{H}^{(r-1)}_r$ denote now the terms of $H^{(r-1)}_r$ not
belonging to ${\cal M}_{res}$. The homological equation at the r-th
step has the form
\begin{equation}\label{homores}
\{i\omega_1^*q_1p_1+i\omega_2^*q_2p_2,\chi_{res,r}\}
=-\tilde{H}^{(r-1)}_r ~~.
\end{equation}
Thus, the equation is diagonal with respect to all monomials
belonging to $\tilde{H}^{(r-1)}_r$, i.e., for every monomial term
$h^{(r-1)}_{k_1,l_1,k_2,l_2}q_1^{k_1}p_1^{l_1}q_2^{k_2}p_2^{l_2}$ in
$\tilde{H}^{(r-1)}_r$ we add a corresponding term in $\chi_{res,r}$
with coefficient
$h^{(r-1)}_{k_1,l_1,k_2,l_2}/(i((k_1-l_1)\omega_1^*+
(k_2-l_2)\omega_2^*))$.

An alternative algorithm to the above would be to introduce a 
local action variable $J_1=I_1-I_1^*$ to be treated as a new small 
parameter. However, in practice we found that this approach has 
worse convergence properties than the ones induced by the normal 
form after the manipulation of the Hamiltonian as in Eq.(\ref{hamtrick}). 
The latter's efficiency can be tested by numerical examples as 
below.

\subsection{Implementation}

\begin{figure}
\centering
\includegraphics[scale=0.35]{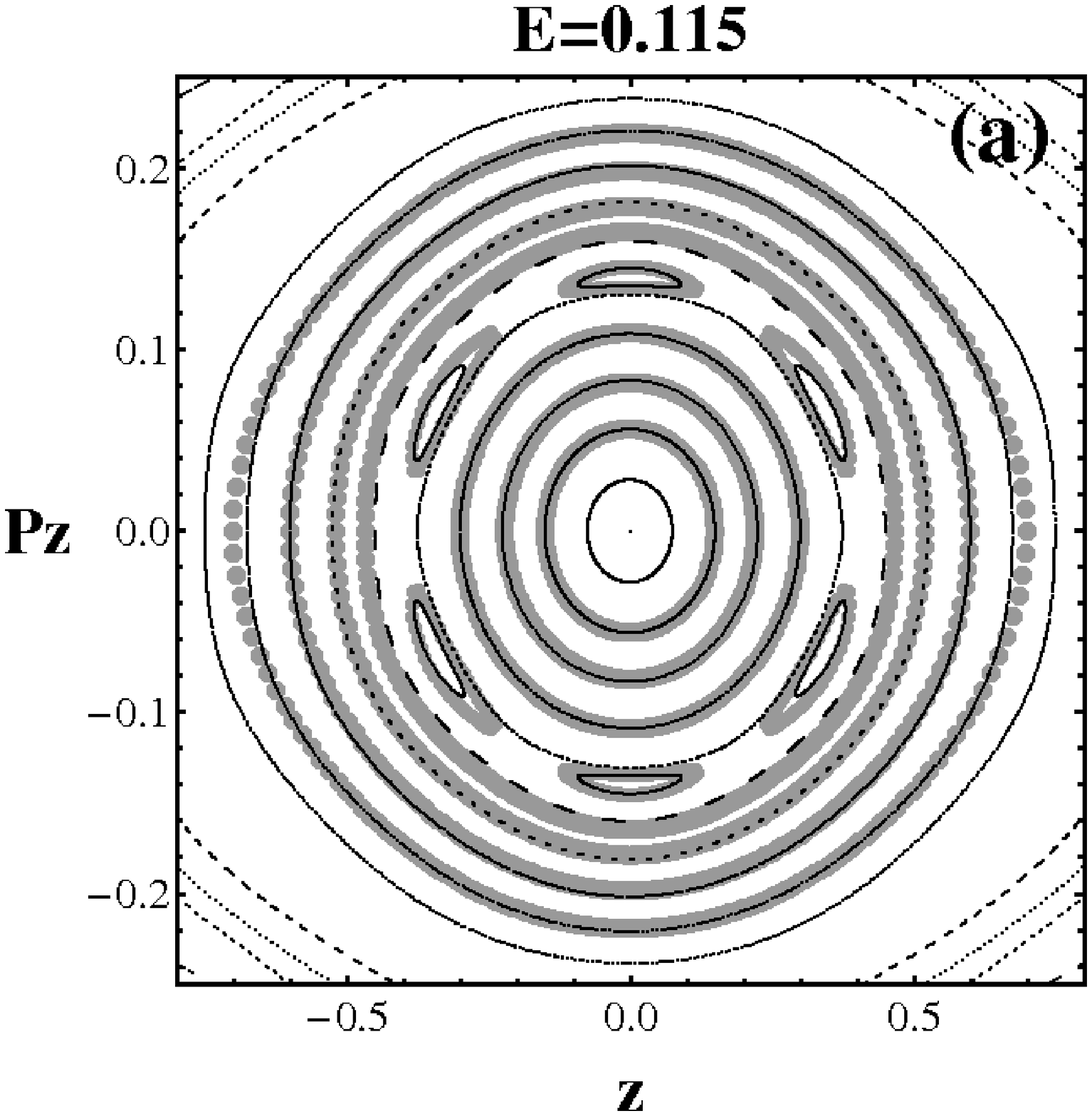}
\includegraphics[scale=0.35]{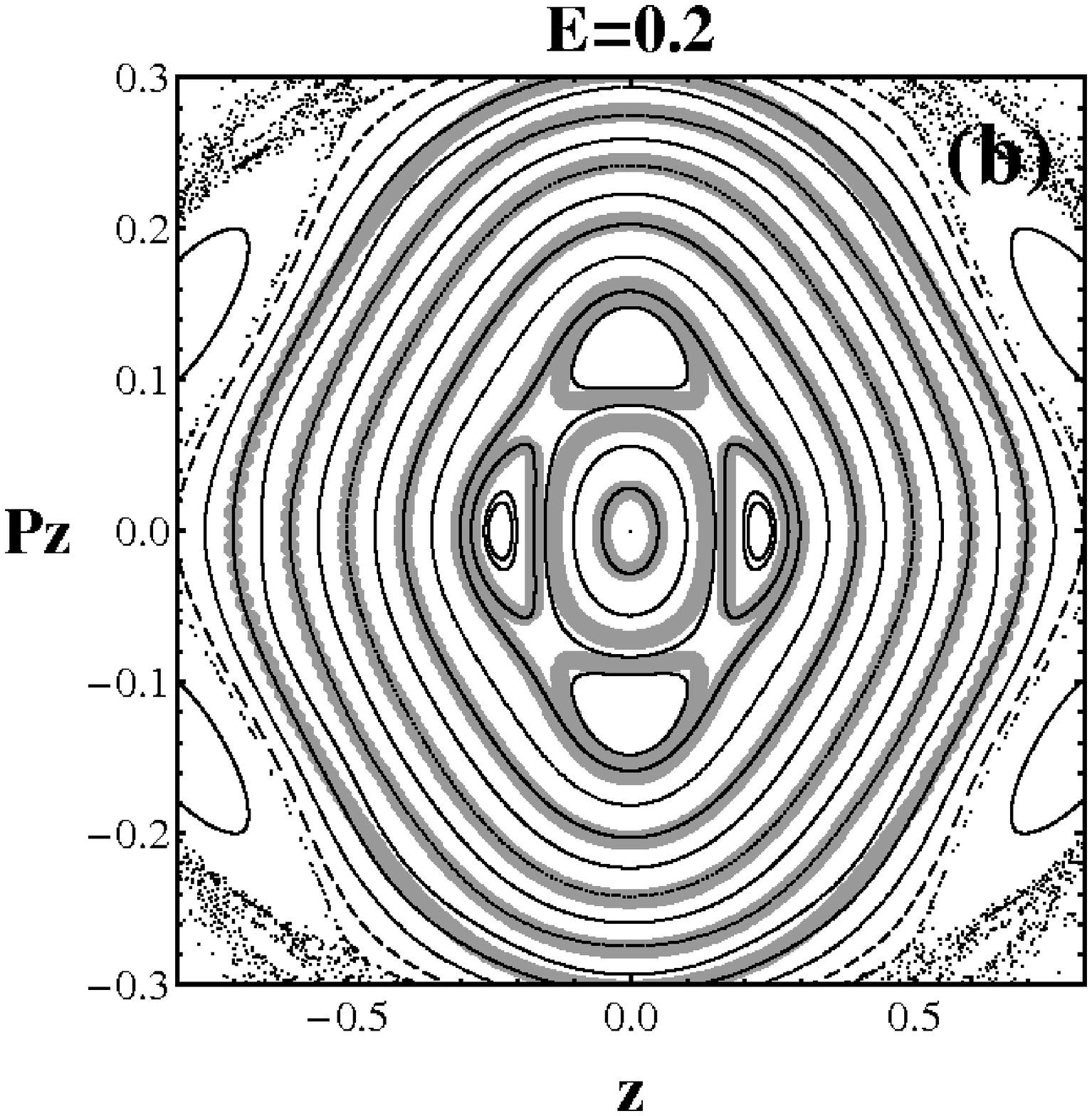}
\caption{\small The Poincar\'{e} surface of section
($z,~p_z,~\rho=0$) (black curves) superimposed with the curves of
equal values of the integral of the resonant normal form (gray
points) for (a) $E=0.115$ and (b) $E=0.20$. The $3:1$ resonant
islands (two sets of 3 islands) in (a) and the $2:1$ resonant
islands (two sets of 2 islands) in (b) are very well reproduced by
the resonant normal form.} \label{resi}
\end{figure}

The resonant formal integral $I_{res}$ can be transformed to an expression
in the original variables in the same way as for the non-resonant case. 
After $r$ normalization steps we have
\begin{equation}\label{phiresold}
\Phi_{res}(q_1,p_1,q_2,p_2)=
\exp(-L_{\chi_{res,1}})\circ\exp(-L_{\chi_{res,2}})\ldots
\circ\exp(-L_{\chi_{res,r}})(im_1q_1p_1+im_2q_2p_2)~~~.
\end{equation}
This allows, again, to compare theoretical with numerical curves on
the Poincar\'{e} surface of section. Figure \ref{resi} shows an
example, referring to two different resonances, namely 3:1
(Fig.\ref{resi}a) at the energy $E=0.115$, and 2:1
(Fig.\ref{resi}b), at the energy $E=0.20$, same as in
Fig.\ref{nonresi}b. In both cases, the energy is taken close to but
above the bifurcation energy value, which, using the normal form
approach (see subsection 4.1) at the 8-th order, is found to be
$E_{1/3}=0.097279$ and $E_{1/2}=0.188036$ (the values determined
numerically are $E_{1/3}=0.097253$ and $E_{1/2}=0.188015$
respectively). In both cases the resonant normal form represents
well the corresponding islands of stability. Furthermore, in both
cases the resonant normal form represents also well the
invariant curves which surround the center both in the interior and
the exterior of the resonant zone. The accuracy of the normal form
computations, which depends on the behavior of the remainder of the
normal form series as a function of the normalization order, is
examined in detail in the next subsection. Here, we emphasize that
the overall limit of validity of the resonant normal form approach
is defined by the appearance of other resonances, of different order
than the resonance under consideration. These extra resonances are
conspicuous in the outer parts of the surface of section of
Figs.\ref{resi}a,b, where we see also the beginning of a chaotic sea
surrounding the main domain of stability.

\subsection{Asymptotic behavior}
\begin{figure}
\centering
\includegraphics[scale=0.45]{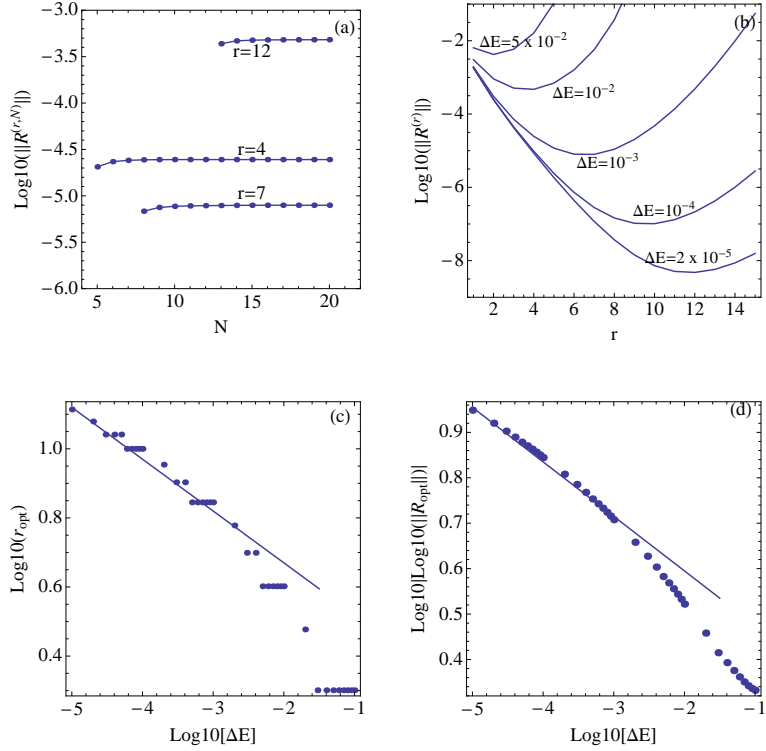}
\caption{\small Asymptotic behavior of the resonant normal form. 
(a-d) Same as in Fig.\ref{nrasym}a-d, but for the resonant normal 
form computation with parameters as indicated in the figure.
The norm definition is given in Eq.(\ref{rnormres}).}
\label{resasym}
\end{figure}

The asymptotic behavior of the resonant normal form is
probed again by numerical experiments, in the same way as in
subsection 4.3 for the non-resonant normal form. Here, since both
frequencies $\omega_1^*$ and $\omega_2*$ are fixed, we introduce a
slight modification of the norm definition with respect to
Eq.(\ref{rnorm}), taking into account also the different choice of
book-keeping rule. Thus, we set:
\begin{eqnarray}\label{rnormres}
||R^{(r,N)}||_{E,\Delta E,\beta} &=&
\sum _{s=r+1}^{N}
\sum_{\begin{array}{c}^{k_1,l_1,k_2,l_2>0}\\
^{k_1+l_1+k_2+l_2=2,4,\ldots 2s+2}\end{array}}
\Bigg(
|G_{s,k_1,l_1,k_2,l_2}^{(r)}|
((E-\Delta E)/\omega_1^*))^{k_1+l_1\over 2}\nonumber\\
&\times&
|\beta|^{k_2}
\left({2\Delta E\over 1+\beta^2(\omega_2^*)^2((E-\Delta E)/\omega_1^*))}
\right)^{k_2+l_2\over 2}\Bigg)~~.
\end{eqnarray}
Figure \ref{resasym}, which is quite similar to Fig.\ref{nrasym},
clearly shows that the resonant normal form series exhibit
asymptotic properties analogous to the non-resonant one.
Nevertheless, a comparison of Figs. \ref{nrasym}b and \ref{resasym}b
shows that the overall error of the resonant formal integrals is
uplifted by about one order of magnitude with respect to the
non-resonant case for similar levels of mirror oscillation energy
$\Delta E$. The exponent found in Fig. 8c is also close to 0.2.
Finally, we observe that the exponential regime for the scaling of
the optimal remainder with $1/\Delta E$ holds for mirror oscillation
energies below a value $\Delta E \leq10^{-3}$.

\subsection{Normal form determination of the threshold to global chaos}
The domain of regular motion around the central periodic orbit shrinks, 
in general, as the energy increases. This domain shrinks to zero at an 
energy $E_t$ where the central orbit exhibits (for the first time) a 
transition from stability to instability. The energy $E_t$ can be found 
numerically by computing the monodromy matrix of the central periodic 
orbit at different energies. Numerically, we find $E_t=E_1\approx$0.36688.

The energy $E_t$ marks the limit of applicability of the normal forms 
in magnetic bottle problem, in either the non-resonant or resonant form. 
We show now how the energy $E_t$ can be determined by normal 
form computations. We proceed as follows: at the energy $E=E_t$,
the two frequencies $\omega_1$ and $\omega_2$ become equal
(Contopoulos, 1968). We then have the bifurcation of two equal
period resonant periodic orbits from the central orbit. Thus we have
$E_t=E_{1/1}$. However, $E_{1/1}$ can be computed as indicated 
in subsection 4.1, for the resonance $m_1=m_2=1$.  

\begin{figure}
\centering
\includegraphics[scale=0.6]{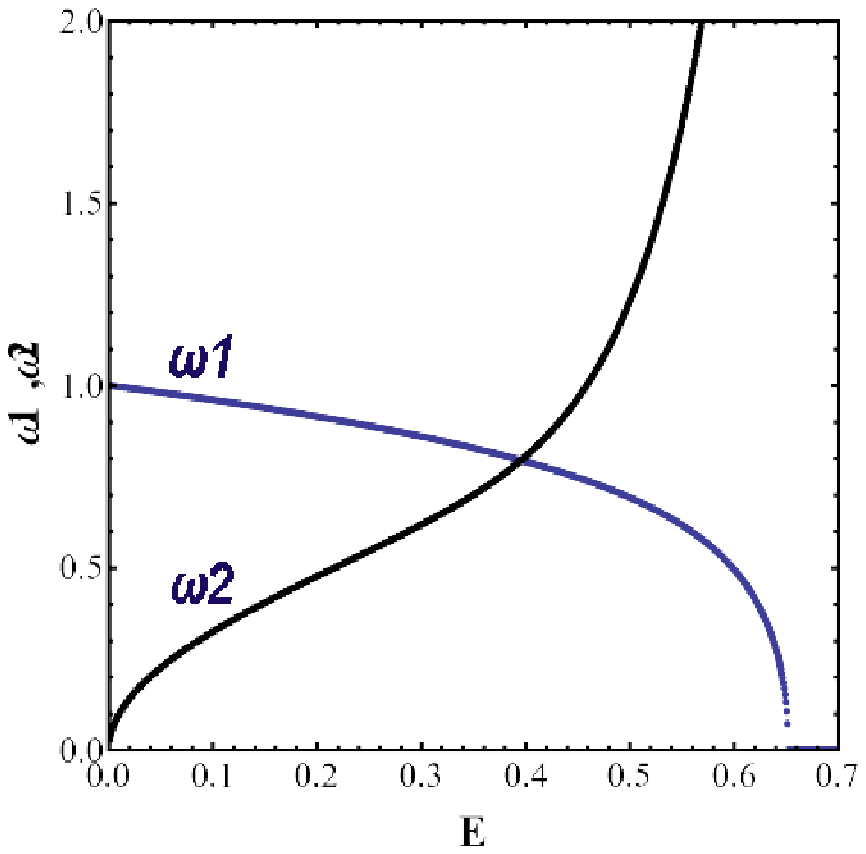}
\includegraphics[scale=0.3]{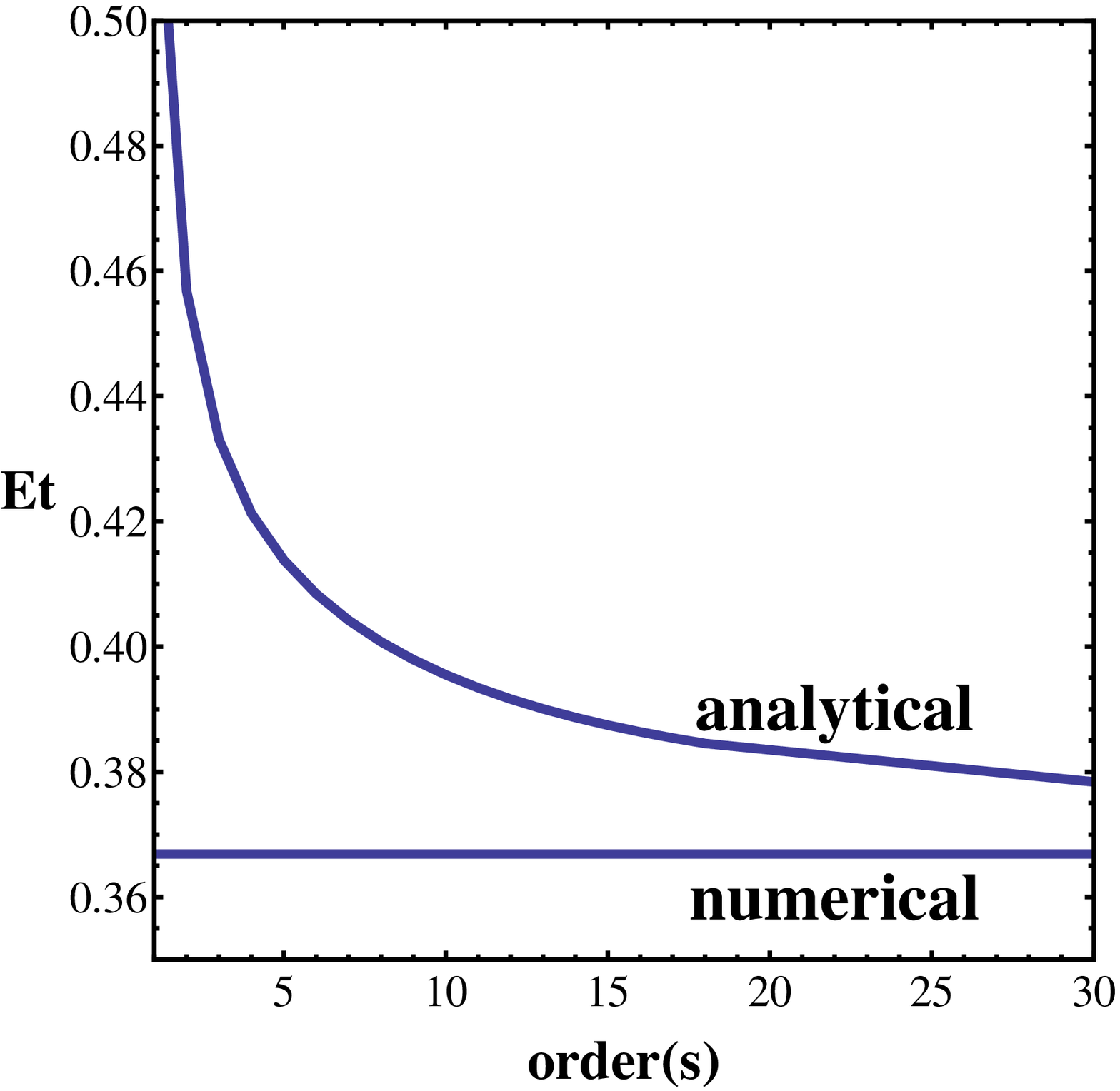}
\caption{\small  (a) Theoretical values for the frequencies $\omega_1$ 
and $\omega_2$ as functions of the energy $E$ by the normal form at the
normalization order $r=10$. The intersection point yields an
estimate of the energy $E_t$ where the first transition of the
central periodic orbit from stability to instability takes place.
(b) The theoretical estimate $E_t$ as a function of the
normalization order $r$. The horizontal line corresponds to the
numerically computed value of $E_t$. } \label{ome11}
\end{figure}

The accuracy of the theoretically computed value depends on the normalization 
order $r$. Fig.\ref{ome11}a shows the frequencies $\omega_1$ and $\omega_2$ 
as functions of the energy $E$ using the normal form at the normalization 
order $r=10$. The intersection point of the two curves yields a theoretical 
estimate $E_t$=0.39550, which has an error $\delta E\approx$0.0286. The error 
is reduced as $r$ increases. Figure \ref{ome11}b shows the estimate for 
$E_t$ as a function of $r$ up to $r=30$. The convergence to the numerically 
computed value is rather slow, the error being about $10^{-2}$ at $r=30$.

\section{Conclusions}
In the present paper we explored the limits of applicability of the
normal form theory in a polynomial magnetic bottle Hamiltonian model. 
We focused on a new algorithm for the construction of the normal 
form and the computation of quasi-integrals (truncated formal integrals) 
in cases of resonance between the gyration and the mirror frequencies. 
Furthermore, we explored the asymptotic behavior of both the non-resonant 
and the resonant normal form series. Our main conclusions can be 
summarized as follows:

1) We explored the asymptotic behavior of the non-resonant normal 
form series. Extending the computations at high normalization series 
confirms the basic theoretical picture that the behavior of the 
normalization is asymptotic. Namely, although the size of
the remainder of the normal form series goes to infinity when the
normalization order $r$ tends to infinity, we observe that 
initially (at low order $r$) the size of the remainder decreases 
with $r$. The remainder becomes minimum at an optimal order $r$ 
which scales approximately as an inverse power of the mirror 
oscillation energy $\Delta E$. The size of the optimal remainder 
is found to be exponentially small in $1/\Delta E$. We estimate 
numerically the exponents related to this asymptotic behavior.

2) The non-resonant normal form allows to compute theoretically 
the energies at which resonant periodic orbits of any resonance 
$m_2:m_1$ (between the mirror and the gyration frequencies, with 
$m_1,m_2$ integers) bifurcate from a `central' (equatorial) orbit. 
We propose a novel computation of the normal form in the case of 
resonances. This is based on combining two algorithmic techniques 
called `detuning' (Pucacco et al. 2008) and `book-keeping' 
(Efthymiopoulos 2012). We give numerical examples of applicability 
of the resonant formal series in the case of the 3:1 and 2:1 
resonances, and demonstrate their ability to predict the form of 
the phase portrait in the neighborhood of each resonance. 

3) We explore the asymptotic behavior also of the resonant formal 
series, which is found to be qualitatively similar to the non-resonant 
case, and estimate numerically the associated exponents.

4) The suggested normal form computations serve to predict two
results regarding the onset of chaos. i) At low energies, one
can estimate the limits of the domain of stability around the
central equatorial orbit, i.e. how far from this orbit (in
phase space) does chaos become important. ii) The onset of
global chaos can be approximated by the energy value where the
central orbit suffers its first transition from stability to
instability, which coincides with the bifurcation energy 
of the 1:1 resonance.

\section*{Acknowledgments}
This research was supported in part by the research committee of
the Academy of Athens (grant 200/815).


\begin{thebibliography}{}
\bibitem{} 
Arnold, V.I., Kozlov, V.V. and Neishtadt, A.I. : 1988, 
"Mathematical Aspects of Classical and Celestial Mechanics", 
Springer, Berlin .
\bibitem{}
Baider, A., and Sanders, J.A.: 1991, J. Diff. Eq. {\bf 92}, 282.
\bibitem{}
Benettin, G. and Sempio, P.: 1994, Nonlinearity {\bf 7}, 281.
\bibitem{}
Benettin, G., Henrard, J. and  Kuskin, S. : 1999, "Hamiltonian
Dynamics. Theory and Applications", Springer, Berlin.
\bibitem{}
Churchill R.C., Pecelli G. and Rod D.L.: 1980, Arch. for
Rat. Mech. and Anal. {\bf 73}, 313.
\bibitem{}
Contopoulos G.:1965, Astrophys. J., {\bf 142}, p.802.
\bibitem{}
Contopoulos G.: 1968, Astrophys. J. {\bf 153}, 83.
\bibitem{}
Contopoulos, G. and Harsoula, M.: 2008, Int. J. Bif. and Chaos, {\bf
18}, 2929
\bibitem{}
Contopoulos, G. and Harsoula, M.: 2010a, Cel. Mech. Dyn. Astron.,
{\bf 107}, 77.
\bibitem{}
Contopoulos G. and Harsoula M.: 2010b, Int. J.Bif. and Chaos {\bf
20}, 2005.
\bibitem{}
Contopoulos, G. and Vlahos, L.: 1975, J. Math. Phys., {\bf 16}, 1469.
\bibitem{}
Contopoulos, G. and Zikides, M.: 1980, Astron. Astroph., {\bf 90}, 198.
\bibitem{}
Cushman, R., and Sanders, J.A.: 1986, in M. Golubitsky and J. Guckenheimer 
(eds), Amer. Math. Soc. Contemporary Mathematics, {\bf 56}, Providence, RI. 
\bibitem{}
Dendy, R.O.: 1993, `Plasma Physics: an introductory course', Cambridge
University Press, Cambridge, UK.
\bibitem{}
Dragt, A. J.: 1965,  Rev. Geophys. Space Phys., {\bf 3}, 255.
\bibitem{}
Dragt, A. J. and Finn, J. M.: 1979,  J. Math. Phys., {\bf 20},
2649.
\bibitem{}
Dunnett, D.A., Laig, E.W. amd Taylor, J.B.: 1968, J. Math. Phys.,
{\bf 9}, 1819.
\bibitem{}
Efthymiopoulos, C.: 2008, Cel. Mech. Dyn. Astron., {\bf 102},  49.
\bibitem{}
Efthymiopoulos, C.: 2012, in Cincotta, P., Giordano, C. and
Rfthymiopoulos, C. (eds) "Chaos, diffusion and non-integrability in
Hamiltonian systems- Applications to Astronomy", Asocation Argentina
de Astronomia Workshop Series, {\bf 3}, p.3-146.
\bibitem{}
Efthymiopoulos, C., Giorgilli, A. and Contopoulos , G.: 2004, J.
Phys. A, {\bf 37}, 10831.
\bibitem{}
Elphick, C.: 1988, J. Phys. Lett. A, {\bf 127}, 418.
\bibitem{}
Engel, U., M., Stegemerten, B. and Eckelt, P.: 1995, J. Phys. A,
{\bf 28}, 1425.
\bibitem{}
Gurnett, D.A., and Bhattacharjee, A.: 2005, `Introduction to Plasma
Physics: With Space and Laboratory Applications', Cambridge University
Press, Cambridge, UK.
\bibitem{}
Harsoula M., Kalapotharakos C. and Contopoulos G.: 2011,
Int. J. Bif. and Chaos, {\bf 21}, 2221.
\bibitem{}
Howard, J.: 1970, Physics of Fluids, {\bf 13}, 2407.
\bibitem{}
Jackson, J,D.: 1962, "Classical Electrodynamics", Wiley , New York.
\bibitem{}
Kruskal M.: 1962, J. Math. Phys. {\bf 3}, 806
\bibitem{}
Lichtenberg A.J. and Lieberman M.A.: 1992, {\it "Regular and
Chaotic Dynamics"}, Springer-Verlag, 1992.
\bibitem{} 
McNamara, B.:1978, J. Math.Phys. {\bf 19}, 2154.
\bibitem{} 
Meyer, K.R.:1984, Funk. Ekvacioj {\bf 27}, 261.
\bibitem{} 
Neishtadt, A.I.: 1981, J. Applied Math. Mech. {\bf 45}, 58.
\bibitem{} 
Neishtadt, A.I.: 1984, J. Applied Math. Mech. {\bf 48}, 197.
\bibitem{} 
Northrop, T.G.: 1964, "The adiabatic motion of charged particles", 
Wiley, London.
\bibitem{}
Pucacco, G., Boccaletti, D. and  Belmonte C.: 2008, Cel. Mech. Dyn. 
Astron., {\bf 102}, 163
\bibitem{} 
Sanders, J.A., Verhulst, F., and Murdock, J.: 2007, `Averaging Methods in 
Nonlinear Dynamical Systems', Springer, Berlin.

\end{thebibliography}
\end{document}